\documentclass[11pt]{emulateapj}\usepackage{times}
\usepackage{amssymb}

\begin{document}

\shorttitle{\textsc{Pulsar model of the high energy phenomenology
of LS 5039}} 
\shortauthors{\textsc{A. Sierpowska-Bartosik \& D. F. Torres}}

\title{\textsc{Pulsar model of the high energy phenomenology
of LS 5039 }}
\author{Agnieszka Sierpowska-Bartosik\altaffilmark{1} \& Diego F. Torres\altaffilmark{2,1}}

\altaffiltext{1}{Institut de Ci\`encies de l'Espai (IEEC-CSIC),
              Campus UAB,  Torre C5, 2a planta,
              08193 Barcelona, Spain.
              Email: agni@ieec.uab.es}
\altaffiltext{2}{Instituci\'o Catalana de Recerca i Estudis Avan\c{c}ats (ICREA).
              Email: dtorres@ieec.uab.es}

\begin{abstract}
Under the assumption that LS 5039 is a system composed by a pulsar rotating around an O6.5V star in a $\sim 3.9$ day orbit,  
we present the results of a theoretical modeling of the high energy phenomenology observed by the High Energy Stereoscopy Array (H.E.S.S.). This model (including detailed account of the system geometry, Klein-Nishina inverse Compton, $\gamma$-$\gamma$ absorption, and cascading)  is able to describe well the rich observed phenomenology
found in the system at all timescales, both flux and spectrum-wise. 
\end{abstract}


\section{Introduction}

A few binaries have been identified as variable very-high-energy (VHE) $\gamma$-ray sources (Aharonian et al. 2005a,b; 2006, Albert et. al. 2006, 2007). For LS 5039, a periodicity in the $\gamma$-ray flux, consistent with the orbital timescale as determined by Casares et al. (2005), was found with amazing precision (Aharonian et al. 2006). Short timescale variability displayed on top of this periodic behavior, both in flux 
and spectrum, was also reported.
Indeed, the H.E.S.S. observations of LS 5039  ($\sim70$ hours distributed over many orbital cycles, Aharonian et al. 2006)
constitute one of the most detailed datasets of 
high energy astrophysics. Accordingly, it has been subject of intense theoretical studies (e.g., Bednarek 2006, 2007; Bosch-Ramon et al. 2005; B\"ottcher 2007; B\"ottcher \& Demer 2005; Dermer \& B\"ottcher 2006; Dubus 2006a,b; Paredes et al. 2006; Khangulyan et al. 2007). However, the rich observed phenomenology of the system has  precluded  an unifying explanation.

The discovery of a jet-like radio structure in LS 5039 and the fact of it being the only
radio/X-ray source co-localized with an EGRET detection, prompted a microquasar interpretation (Paredes et al. 2001). However, the current
findings at radio and VHE $\gamma$-rays in the cases of LS I +61 303 (Dhawan et al. 2006, Albert et al. 2006) or PSR B1259-63 (Aharonian et al. 2005), gave the
perspective that all three systems are different realizations of the same
scenario: a pulsar-massive star binary. Dubus (2006a,b) has studied these similarities. He provided simulations of the extended radio emission in the case of LS 5039, showing that the features found in high resolution radio observations could also be interpreted as the result of a pulsar wind. This result, the rest of similarities found with systems for which a pulsar companion is known, and the assessment of the low X-ray state (Martocchia et al. 2005), makes the pulsar hypothesis tenable. High energy emission from pulsar binaries have been subject of study for a long time (e.g., Moskalenko et al. 1993, Moskalenko 1995; Arons \& Tavani 1993; Maraschi \& Treves 1981; Romero et al. 2001). 
But the dichotomy between pulsars or black hole binaries is not settled (see, e.g., Mirabel 2006). Here, for testing whether the H.E.S.S. observations could be explained in a pulsar scenario, we have assumed that the compact object in LS 5039 is in fact a pulsar. Under this assumption, 
this Letter presents the results of a theoretical modeling of LS 5039 which is able to describe reasonably well the phenomenology found in the system.

\section{The model}

Following Sierpowska \& Bednarek (2005) (see also Bednarek 1997), we consider that the volume 
of the binary system is separated by a shock wave into two regions with different
properties, with the shock appearing as a result of the collision between the pulsar and the massive star winds. The position of the shock is defined by the parameter $\eta =  L_{\rm SD} /(\dot{M} V_{\rm w}c)$, where the wind velocity, $V_{\rm w}$, depends on the radial distance from the 
massive star, $\dot{M}$ is the star mass-loss rate, and $L_{\rm SD} $ is the spin-down luminosity of the pulsar (Ball \& Dodd, 2001). Note that for $\eta < 1$ the star wind dominates over the pulsar's: the termination shock 
wraps around it. It is assumed that the pulsar and stellar winds
are symmetric. In the case of LS 5039 (see Table 1) the value of $\eta$ is between $0.5$ (periastron) and $0.3$ (apastron).
Between the pulsar and the shock, inside the pulsar wind zone (PWZ), we assume that relativistic 
leptons are frozen in the magnetized pulsar wind which propagate 
radially from the pulsar. Therefore, their synchrotron losses are neglected in our cascade calculations. We consider then (Klein Nishina) inverse Compton (IC) cascades in this PWZ, assuming that the thermal radiation from the massive star dominates in this region  (see e.g., Ball \& Kirk 2000). Synchrotron emission are typically considered in models where the $\gamma$-emission mainly comes from the shock region, especially applicable in the case of binaries with more eccentric orbits (e.g., Kirk, Ball \& Skjaeraasen 1999). In such very eccentric massive binaries with large separations the radiation field of the massive star does not dominate over the whole orbit and IC scattering is not effective in the PWZ. 
We assume that the cascade initiated by a primary lepton in the
PWZ develops in one dimension, i.e. in the direction of propagation of the primary 
particle. 
The charged products of this cascade arrive finally to the shock region in the 
pulsar wind and follow the flow along the shock surface. 
 Due to the magnitude of the opacity to IC and $\gamma$-$\gamma$, most electrons effectively interact in the PWZ. Indeed, for instance for inclination $i=30^o$, for 1 TeV (100 GeV) electrons, the opacity to IC is $\sim1.5$ $(\sim 9)$ in the PWZ. The energy carried by electrons that actually reach the shock to be trapped by the magnetic field is less than $\sim10\%$ of the primary injected energy.
The secondary $\gamma$-rays move into the massive star wind region (MSWR). 
Some of them escape the binary system but a significant part can be absorbed due to $\gamma$-$\gamma$ process. These cascades are followed by means of a Monte Carlo procedure (explained in general terms in Sierpowska \& Bednarek 2005). The details of the LS 5039 implementation worked out for this paper (including the detailed account of the system geometry, the computation of opacities to the different processes as a function of orbital phase, and many other details specific to our study) will be presented elsewhere. 
We assume that the pulsar-provided injected power in relativistic electrons is a fraction of the spin-down power, and that they are described by a power-law in energy. 
Table 1 presents the few free parameters of the model.\footnote{The first two free parameters shown are obviously related. However, note that
the spin-down power defines the position of the shock in the system, through $\eta$, which also relates the stellar mass-loss rate $\dot M$ and $L_{\rm SD}$.}
Additional model parameters are needed, but these are bound to fixed values (given also in Table 1) under multiwavelength observations  (see, e.g., the papers quoted at the introduction).

\section{Results}

\begin{figure*}
\includegraphics[width=.5\textwidth]{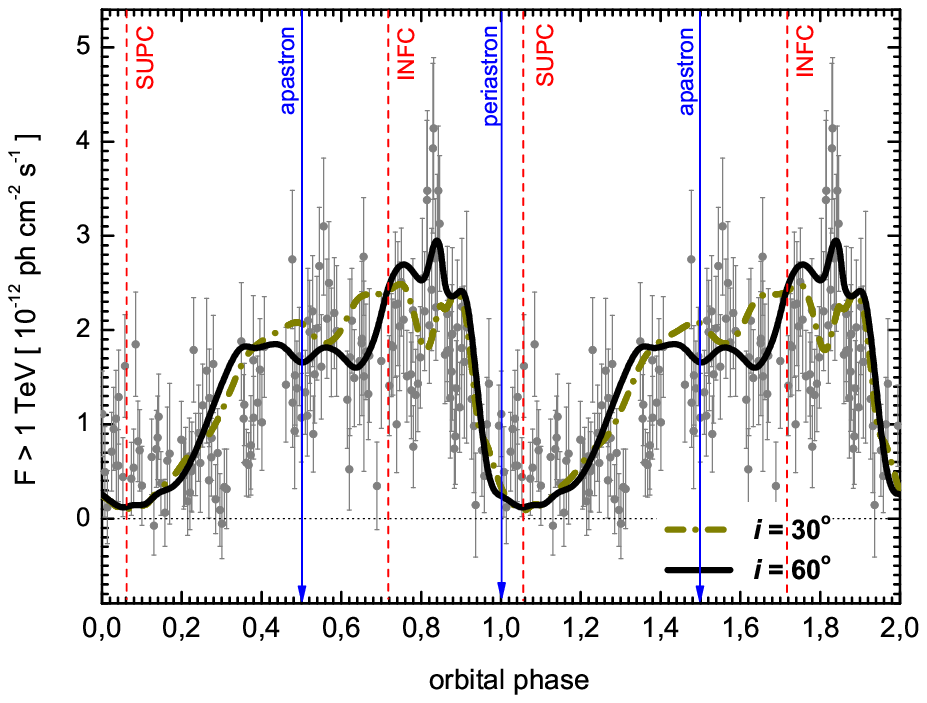}
 \includegraphics[width=.5\textwidth]{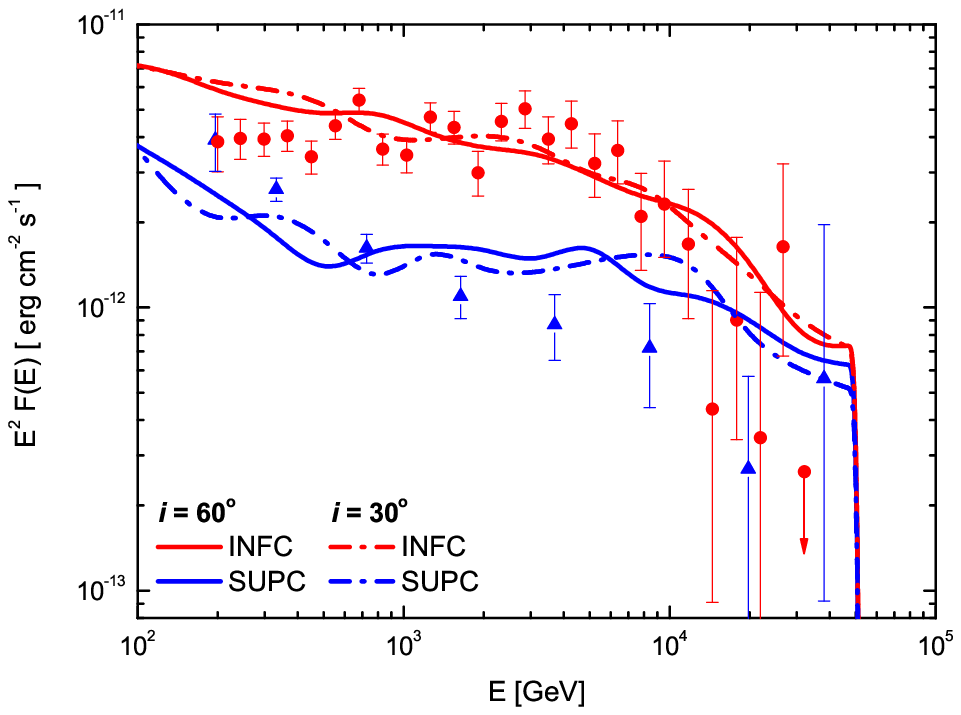}
\caption{Left: H.E.S.S. run-by-run folded observations of LS 5039 with the results of the theoretical model for two different inclination angles.  Right:  H.E.S.S. VHE spectra around INFC and SUPC together with theoretical predictions in equal phase intervals. }
\label{LC}
\end{figure*}

Figure \ref{LC} shows the phase folded H.E.S.S. data for the integral flux above 1 TeV (Aharonian et al. 2006) and the results of our theoretical model. Casares et al.'s (2005) ephemeris was used to fold the observational data. Each experimental point represents a typical time span of observation of 28 min.
The corresponding phases for apastron, periastron, inferior (INFC), and superior  (SUPC) conjunction are marked. Two periods and two  different inclinations angles are shown. 
The trends in the H.E.S.S. datapoints are reproduced: orbital periodicity, a non-zero emission close to periastron, and the maximum of the fluxes is around INFC.

VHE gamma-rays produced close to the star
suffer severe absorption via pair production. 
The absorption process strongly depends on the direction of propagation with respect to the massive star. Indeed, the cross-section for pair production depends on the angle $\theta$ between the 
VHE gamma-ray and optical photons, and has an energy threshold 
$\propto 1/(1- \cos \theta)$. The level of absorption therefore depends on geometry and leads to orbital modulation of the VHE gamma-ray flux.
For LS 5039, apastron and periastron are near 
INFC and SUPC. Absorption provides flux maxima at INFC, when $\gamma$-photons propagate towards the observer outward the system, and opacities are the lowest (cos 
$\theta \rightarrow 1 )$. Flux minima happen at SUPC, when photons propagate toward the massive star (absorption more effective, opacities the highest), 
as earlier discussed, e.g., by Dubus (2006a).
Absorption alone would however produce strict modulation (zero flux) in the energy range 0.2 to 2 TeV whereas the observations show that the flux at $\sim 0.2$ TeV is stable. Additional processes, i.e. cascading, must be considered to explain the spectral modulation. Our model have these processes consistently included and the lightcurve details arise then as the interplay of the absorption of $\gamma$-rays with the cascading process, in the framework of a varying geometry along the orbit of the system.
It can be seen that the differences between the two lightcurves corresponding to different inclinations are not large, but it is the detail at  INFC what is worth noticing. 

\begin{figure*}
\includegraphics[width=.5\textwidth]{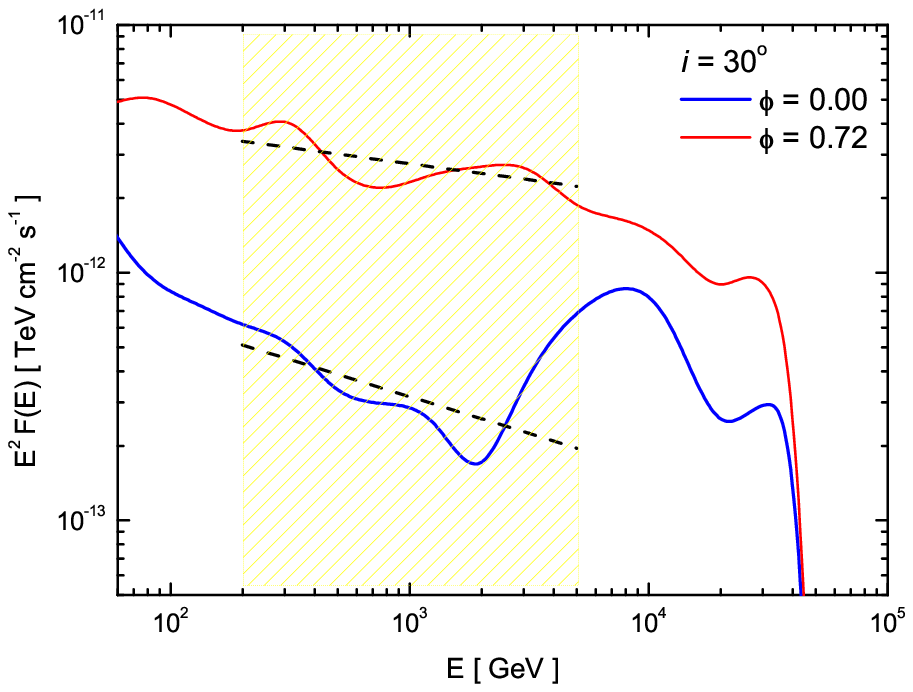}
 \includegraphics[width=.5\textwidth]{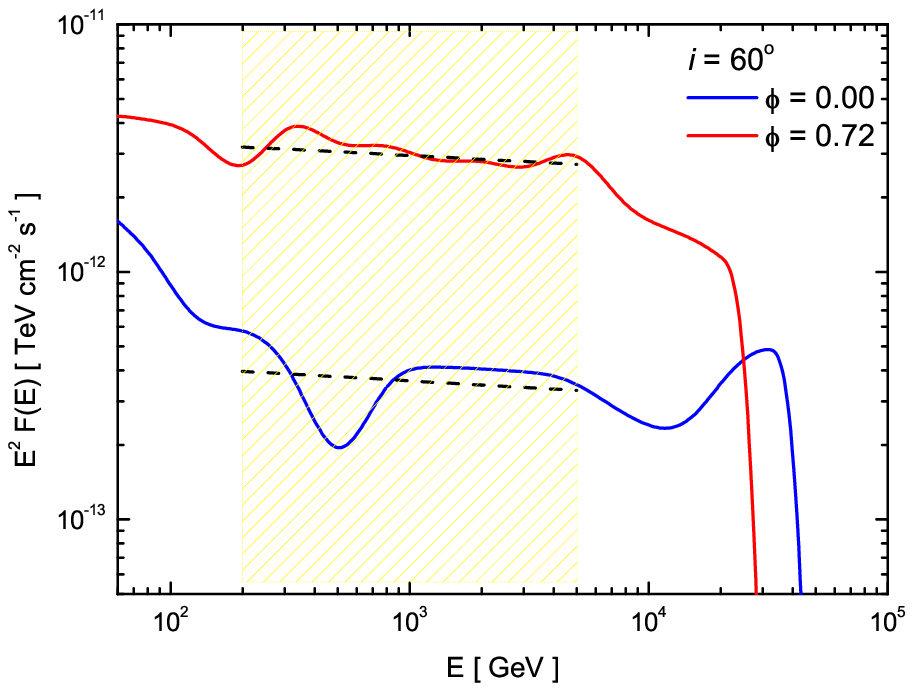}
\caption{Examples of the theoretical results, for different inclinations, for the spectrum of single phases fitted with a power law in the range 0.2--5 TeV. 
}
\label{single-spec}
\end{figure*}
\begin{figure*}
\includegraphics[width=.5\textwidth]{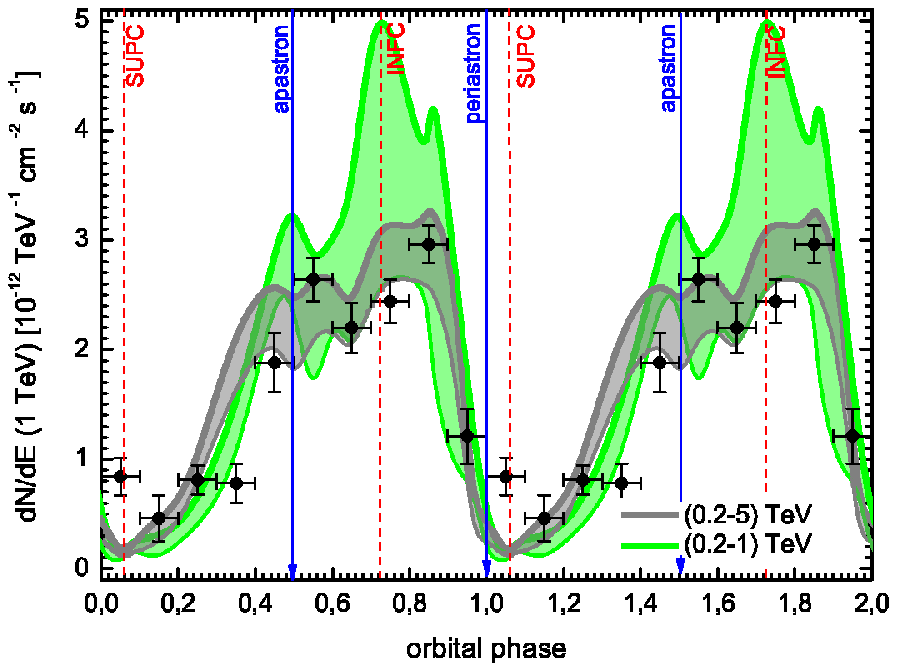}
\includegraphics[width=.5\textwidth]{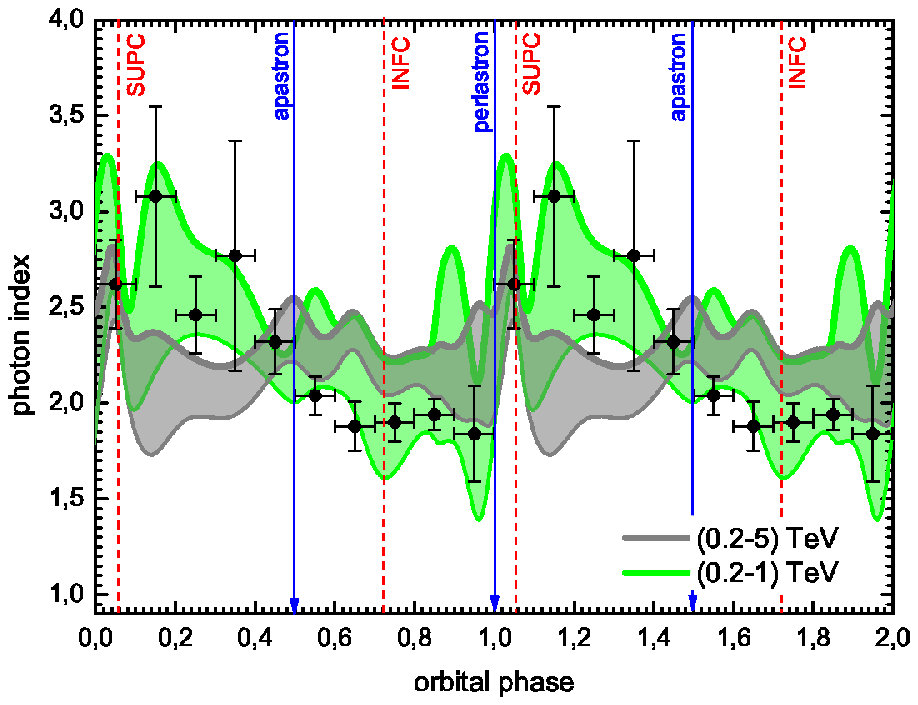}
\includegraphics[width=.5\textwidth]{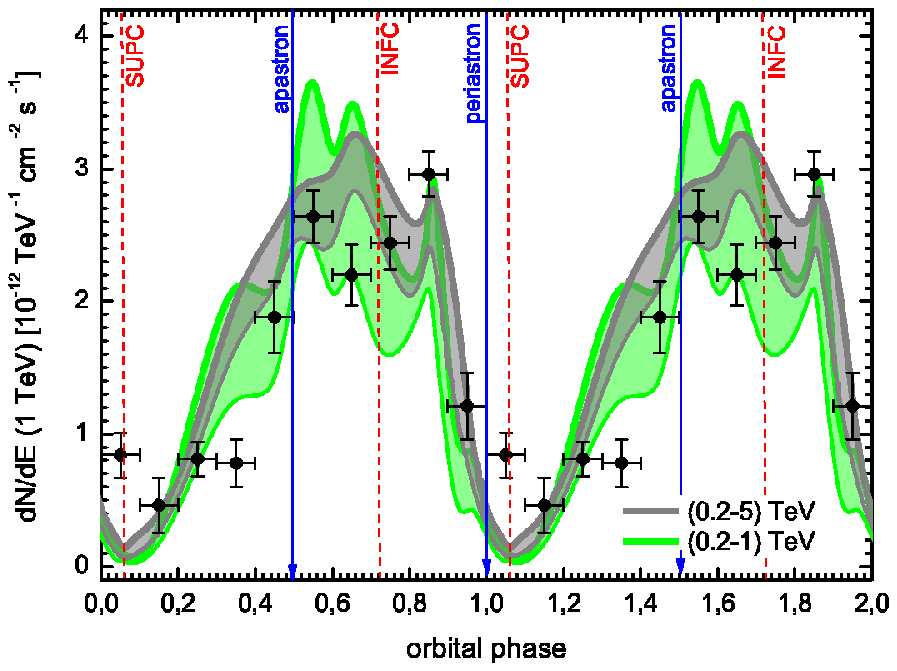}
\includegraphics[width=.5\textwidth]{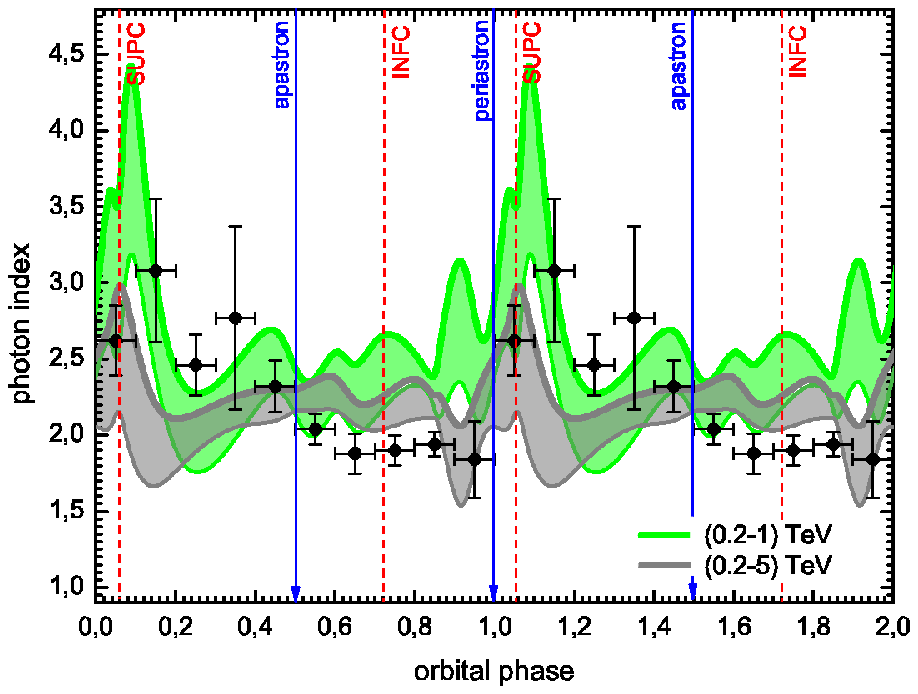}
\caption{Parameters of power-law fittings to the spectra as a function of phase both for the observational H.E.S.S data and our theoretical curves. We show two sets of results corresponding to fitting in different energy ranges (0.2--1 TeV, 0.2--5 TeV), and inclination angles (top: $i=60^0$, bottom: $i=30^0$). The size of the shading represent the error in the parameter fitting of our model.}
\label{phase-bin}
\end{figure*}

The spectra for LS 5039 was presented in two broad orbital phase intervals around INFC and SUPC (Aharonian et al. 2006). 
A clear tendency for a harder spectrum at higher flux level was found. This is reproduced by the theoretical model herein presented. 
In order to compare with the broad phase-intervals spectra presented by H.E.S.S., we have also computed the phase-average of our predictions, see Figure 1. We found an overall agreement.
%
%
%
%
%
H.E.S.S. observations have also provided the evolution of the normalization and slope of a power-law fit to the 0.2--5 TeV data in 0.1 phase-binning (Aharonian et al. 2006). The variability of the spectral index and normalization of these fits along the orbit is impressive.  The use of a power-law fit has to do with low statistics in such shorter sub-orbital intervals: higher-order functional fittings (such as a power-law with exponential cutoff) provide a no better fit and were not justified.  We can directly compare with these H.E.S.S. results fitting our spectral predictions in the same way. 
However, we note that particularly for some phases close to periastron, the theoretical spectrum is badly mimicked by a power-law (see the examples of Figure \ref{single-spec}).  
Using these fits to each of the theoretically predicted spectra as a function of phase along the orbit, we compare with H.E.S.S. observations in Figure \ref{phase-bin}. The overall similarity of the theoretical and observational fittings is notable. Depending on inclination 
there are a few points around INFC for which the observational data seem to be described by a harder spectrum than what we obtain as a result of fitting the theoretical predictions. This can be modeled with a slightly harder injection spectrum for electrons. The normalization data point close to periastron is missed by our predicted fitting ranges. This is the result of our spectrum being badly fitted by a power-law at these phases.

\section{Concluding remarks}

A detailed modeling of LS 5039, under a pulsar scenario, accounting for the system geometry, Klein-Nishina IC, $\gamma$-$\gamma$ absorption, and a Monte Carlo computation of cascading,  describes well the phenomenology
found in the system at all timescales, both flux and spectrum-wise. The predictive power of the model is impressive: it is based on just a handful of free parameters. The results shown here were not obtained varying these parameters searching for a best fit, but rather exploring the predictions under minimal assumptions, what enhances their value.
The overall agreement found does not however constitute
a proof that the system is indeed formed with a pulsar companion (although contributes to circumstantial evidence in this direction). This agreement does provide a framework in which the pulsar assumption can be precisely tested with future observations.

We acknowledge W. Bednarek, J. Cortina, I. Ribas, J. Rico, and N. Sidro for comments, extended use of IEEC-CSIC parallel computers cluster, and support by grants
AYA 2006-00530 and CSIC-PIE 200750I029.

\begin{deluxetable}{lll}
\tablewidth{\textwidth}
\tabletypesize{\tiny}
\tablecaption{Model Parameters}
\tablehead{\colhead{Meaning} & \colhead{Symbol} & \colhead{Adopted value}}  
\startdata
{\bf Free model parameters }\\
\hline 
 Spin-down power of assumed pulsar & $L_{\rm SD}$ & $10^{37}$ erg s$^{-1}$\\
 Fraction of spin-down power injected in relativistic electrons & $\beta$ & $10^{-2}$ \\  
 Slope of the injected power-law description of the electron spectrum & $\Gamma$ & $-2.0$ \\
 VHE cutoff of the injection spectra (consistent with upper end of $\gamma$-spectrum in H.E.S.S. results) & \ldots & $50$ TeV \\
 Inclination of the orbit of the system with respect to the line of sight (consistent with Casares et al. 2005) & $i$ & 30$^0$, 60$^0$\\ 
 \hline
{\bf Most important additional parameters, fixed by multiwavelength observations}\\
 \hline
Radius of star &  $R_\star$ & $9.3\, R_\odot$\\
Mass of star &  $M_\star$ & $23\, M_\odot$\\
Temperature of star & $T_\star$ & $3.9 \times 10^4$ K\\
Mass loss rate of star & $\dot{M}$ & $10^{-7}\, M_\odot\, \rm yr^{-1}$  \\
Stellar wind (SW) termination velocity & $V_{\infty}$ & $2400 \,\rm km\, s^{-1}$ \\
SW initial velocity (SW velocity at radius $r$ from the star: $V_w(r) = V_0+(V_{\infty}-V_0)(1-R_\star/r)^{1.5}$)& $V_0$ & $4\,\rm km\, s^{-1}$ \\
Distance to the system & $D$ & $ 2.5 \,\rm kpc$\\
Eccentricity of the orbit & $\varepsilon$ & $0.35$ \\
Semimajor axis & $a$ & $0.15\, \rm AU$ $\sim 3.5\, R_\star$  \\
Longitude of periastron & $\omega_{p}$ & $226^o$ \\
\vspace{-2mm}
\enddata
\end{deluxetable}

\end{document}